\documentclass[12pt]{iopart}

\usepackage{iopams}
\usepackage{setstack} 
\usepackage{graphics,graphicx,dcolumn,bm,fleqn,epic,eepic,float,epsfig}
\usepackage{amssymb,multirow,rotate,color,float}
\usepackage{epstopdf,times,color,soul}
\usepackage{wrapfig,subfigure,cancel}  
\definecolor{red}{rgb}{1,0,0}
\definecolor{green}{rgb}{0,1,0}
\definecolor{blue}{rgb}{0,0,1}

\definecolor{olivegreen}{rgb}{0.3,0.55,0.1}

\begin{document}

\title[]{
Power grid stability under perturbation of single nodes: Effects of heterogeneity and
 internal nodes}

\author{Matthias Wolff, Pedro G.~Lind and Philipp Maass}

\address{Fachbereich Physik, Universit\"at Osnabr\"uck, Barbarastrasse
  7, 49076 Osnabr\"uck, Germany} \ead{mawolff@uos.de, pelind@uos.de, maass@uos.de}
\vspace{10pt}

\begin{abstract}
Non-linear equations describing the time evolution of frequencies and
voltages in power grids exhibit fixed points of stable grid operation.
The dynamical behaviour after perturbations around these fixed points
can be used to characterise the stability of the grid. We 
investigate both probabilities of return to a fixed point 
and times needed for this return after perturbation of single nodes.
Our analysis is based
on an IEEE test grid and the second-order swing equations for voltage
phase angles $\theta_j$ at nodes $j$ in the synchronous machine
model. The perturbations cover all possible changes $\Delta\theta$ of
voltage angles and a wide range of frequency deviations in a range
$\Delta f=\pm1$~Hz around the common frequency $\omega=2\pi
f=\dot\theta_j$ in a synchronous fixed point state.  Extensive
numerical calculations are carried out to determine, for all node
pairs $(j,k)$, the return times $t_{jk}(\Delta\theta,\Delta \omega)$
of node $k$ after a perturbation of node $j$.  We find that for strong
perturbations of some nodes, the grid does not return to its
synchronous state. If returning to the fixed point, the times needed for the return are
strongly different for different disturbed nodes and can reach values
up to 20 seconds and more.  
When homogenising transmission line and node
properties, the grid always returns to a synchronous state for the
considered perturbations, and the longest return times have a value 
of about 4 seconds for all nodes.
The neglect of reactances between points of power generation
(internal nodes) and injection (terminal nodes) leads to an
underestimation of return probabilities. 

\end{abstract}



\section{Introduction}
\label{sec:intro}

Due to the increasing share of renewable energy sources on power
production, questions concerning the limits, quantification and
control of power grid stability pose new challenges.  These challenges
can be taken on by a combination of methods developed in the fields of
non-linear dynamics, network theory and stochastic modelling
\cite{Backhaus/Chertkov:2013, Timme/etal:2015, Auer/etal:2016, Mureddu/etal:2016, Schiel/etal:2017, Haehne/etal:2018, Tamrakar/etal:2018, Schaefer/etal:2018}.
Generally, for studying power grids, models are needed for the
dynamics of voltages and frequencies as well as for the grid
representation. However, it is still unclear, which detailedness of
modelling is needed for a reliable quantification of grid stability.

\begin{figure*}[t!]
\centering
\includegraphics[width=0.95\textwidth]{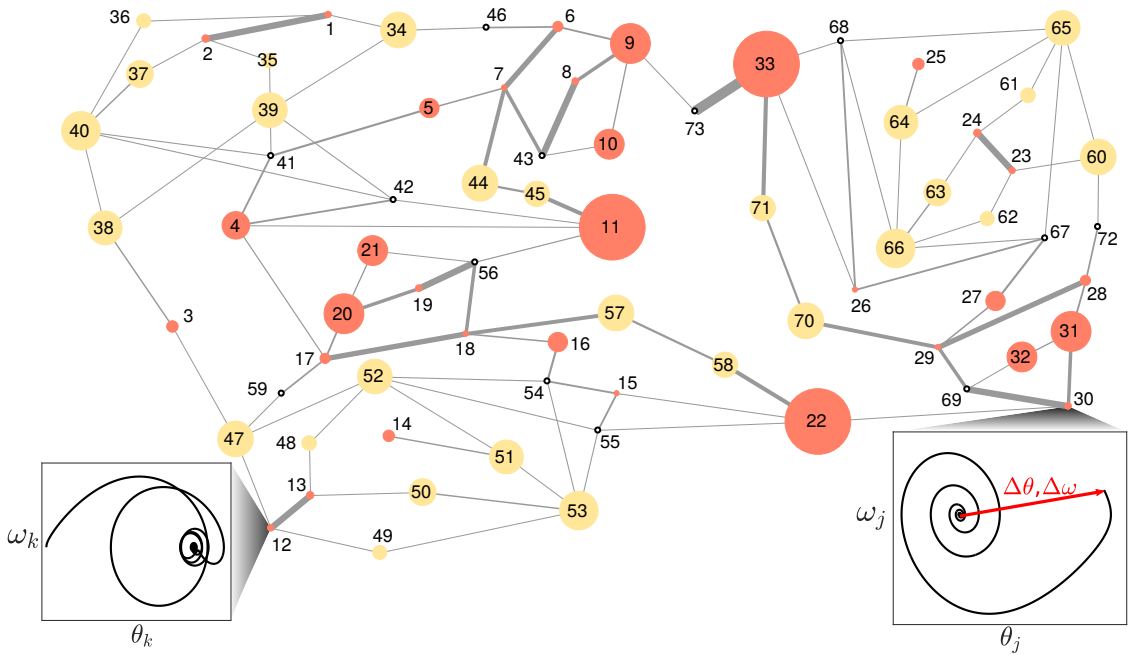}
\caption{\protect Sketch of the IEEE RTS-96 grid, consisting of 33
  generator nodes labelled from 1 to 33 (full red circles), and 40 load nodes (full yellow/open white circles) labelled from
  34 to 73.  The open white circles refer to load nodes with vanishing mechanical power
  and the size of the full symbols has been scaled proportional to their
  mechanical power (for details, see section~\ref{subsec:parameters}).
  The nodes are connected by
  108 transmission lines, where the thickness of the lines mark the
  strength (modulus) of the respective complex admittances, which
   is defined by $[|(Y_0)_{jk}|+|(Y_0)_{kj}|]/2$, where
    $(Y_0)_{jk}$ are the elements of the admittance matrix, see
    section~\ref{subsec:grid-structure}.  In a fixed point state of the
  power flow dynamics, the frequencies at all nodes are equal, 
  corresponding to a state of synchronous operation, where
  the differences $(\theta_j-\theta_k)$ between voltage phase angles
  at all node pairs $(j,k)$ remain constant. After a perturbation of
  node $j$ belonging to the attraction basin of the fixed point, this,
  as well as other nodes $k$, return to their stationary dynamics in the fixed
  point state. A corresponding perturbation of a node
  $j$ is indicated by the red arrow, together with the subsequent
  trajectories of the phase angles and frequencies at this and another
  node $k$.}
\label{fig:grid-illus}
\end{figure*}

For studying power grid dynamics, various approaches have
been used in the literature \cite{Nishikawa/Motter:2015}, which
essentially differ in the treatment of the coupling of generators and
loads to the grid, and the treatment of the voltage dynamics. A widely
used approach is the synchronous machine model, where the dynamics of
both generators and loads are described by the swing equation for
synchronous machines, and where the time-dependence of the moduli of
their complex voltage amplitudes is considered to be
negligible. Refined models with dynamics of voltage amplitude moduli
were considered \cite{Auer/etal:2016, Schmietendorf/etal:2014} and recent 
results indicate that these dynamics do
not affect much the grid stability under stochastic power input
\cite{Schmietendorf/etal:2017}. With respect to the grid representation, often
simplifications are made for specific purposes, e.g.\ by using artificial grid
structures, and/or by taking real network graph
structures, but neglecting heterogeneities in properties of
generators, loads and transmission lines 
\cite{Tamrakar/etal:2018, Filatrella/etal:2008, Lozano/etal:2012, Menck/etal:2014}. 
A further simplification
frequently made is the neglect of reactances between points of power
production/consumption (internal nodes) and power injection/ejection
(terminal nodes).  These reactances can, however, be important,
because they change the effective grid structure from a locally
connected to a fully connected one.

In this work we are investigating power grid response to single-node
perturbations by taking into account strong heterogeneities in line
and node properties typically encountered at high-voltage transmission
grid levels.  For a faithful representation of these heterogeneities,
we use the IEEE RTS-96 grid (reliability test system 96)
\cite{Grigg/etal:1999}, which is sketched in figure~\ref{fig:grid-illus}
together with the methodology used in this work. After perturbation of
a node $j$ in the grid, i.e.\ changes $\Delta\theta$ and $\Delta f$ (or $ \Delta \omega = 2 \pi \Delta f $) of
its voltage phase angle and frequency from their values in a stable
fixed point of synchronous grid operation, the subsequent responses of
the phase angles and frequencies of all nodes are monitored. The range
of considered perturbations covers all possible
$\Delta\theta\in[-\pi,\pi[$, and frequencies in an interval $\pm1$~Hz
with respect to the synchronous frequency, i.e.~$\Delta f\in[-1~{\rm
Hz},1~{\rm Hz}]$. This frequency range is quite wide in
view of the range of primary control of up to 0.4~Hz.\footnote{Even in the 2006 European
blackout, the largest frequency deviations were of the order of 1~Hz
\cite{Li/etal:2007}.}   As for
quantifying grid stability against single-node perturbations, this
can be defined by the probability of return to a fixed point of
synchronous operation for the considered perturbation range, called
basin stability \cite{Menck/etal:2014, Menck/etal:2013}.  In
practice, it matters also how long the grid needs to return to the
vicinity of the synchronous state. We therefore characterise the grid
stability not only by the probability of return but also by the return
times for the perturbations belonging to the attraction basin of the
fixed point. A similar concept has been recently
introduced in \cite{Mitra/etal:2017}.

In addition to getting insight into the basin stability and
distribution of return times of a test grid reflecting real
properties, a further goal of this work is to evaluate the relevance
of the heterogeneities in transmission line and node features for
estimating grid stability, e.g.\ for identifying possible
trouble-causing nodes. To this end we compare our results to those
obtained for simplified, homogenised grid properties, as they are
sometimes used in case studies of grid stability.  We furthermore
investigate effects of reactances between internal and terminal nodes
by using values derived from a method suggested in
\cite{Nishikawa/Motter:2015}. 

The paper is organised as follows. In section \ref{sec:methods} we
describe the modelling of the power grid structure
(section~\ref{subsec:grid-structure}) and of the power flow dynamics
(section~\ref{subsec:dynamics}), and the parameters for simplified and
extended variants of the IEEE RTS-96 grid (section~\ref{subsec:parameters}).
In section~\ref{sec:basin-stability} we investigate the probabilities of return to
the fixed point state of synchronous operation based on the basin stability
measure introduced in \cite{Menck/etal:2013}. The return times are studied
in section~\ref{sec:return-times} as an additional measure 
to characterise the grid response for 
perturbations in the attraction basin of a fixed point. 
For both the basin stability and return times we analyse the changes
found upon homogenising node and line properties
and/or when neglecting internal nodes.  
Our main results are summarised
in section \ref{sec:conclusions} together with a discussion of their
consequences and impact for further studies.

\section{Grid structure modelling and power flow dynamics}
\label{sec:methods}

\subsection{Grid structure modelling}
\label{subsec:grid-structure}

\begin{figure*}[t!]
\centering
\includegraphics[width = 0.9\textwidth]{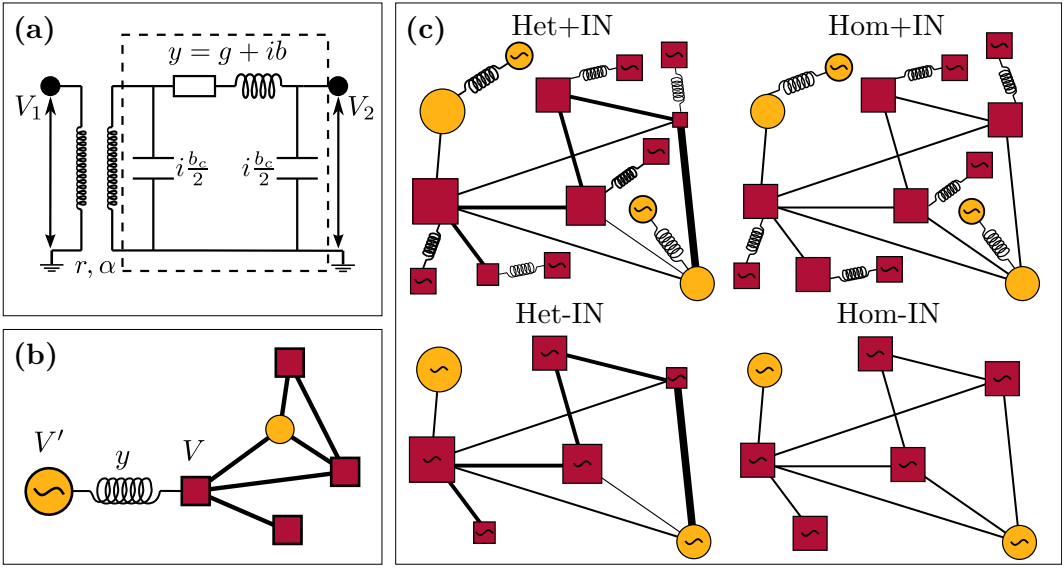}%
\caption{\protect (a) Illustration of the branch model and $\Pi$-model
  (part enclosed in dotted lines) for transformer and transmission
  lines, respectively. (b) Illustration of an internal node (voltage $V'$) and a terminal
  node (voltage $V$), together with their connecting admittance $y$. (c) Four different
  representations of the grid, where "Het" stands for the grid with
  heterogeneities in transmission line and node properties and "Hom"
  for homogenized transmission line and node properties
  (cf.\ section~\ref{subsec:parameters}).  The $\pm{\rm IN}$ indicates
  whether internal nodes are taken into account or not. In (b) and (c) nodes with
  a tilde would act as active nodes in the dynamical equation and can not be eliminated by
  Kron reduction, see section~\ref{subsec:grid-structure}.}
\label{fig02}
\end{figure*}

A common representation of transmission lines in a power grid is given
by the so-called \mbox{$\Pi$-model} \cite{Machowski/etal:2008}, which is illustrated in
figure~\ref{fig02}(a) for a line between two nodes 1 and 2 (part
enclosed by dashed lines).  It consists of an admittance 
$y=g+ib$, with conductance $g$ and susceptance $b$, and
a line charging susceptance $b_{\rm c}$ represented by two capacitors
in parallel to it.  For representing a transformer with turn ratio $r$
and phase shifting $\alpha$, the model can be extended to the branch
model, given by all elements shown in figure~\ref{fig02}(a). For this
branch model, the (complex) outgoing currents $I_{1\to}$ and
$I_{2\to}$ leaving nodes 1 and 2 are linearly related to the (complex)
node voltages $V_1$ and $V_2$,
\begin{equation}
\left[\begin{array}{c}
I_{1\to} \\
I_{2\to}
\end{array}
\right]= 
\left[
\begin{array}{cc}
y_{11} & y_{12} \\
y_{21} & y_{22}
\end{array}
\right] 
\left [
\begin{array}{c}
V_1\\
V_2
\end{array}
\right ] \,,
\label{eq:branch-model}
\end{equation}
where $y_{11}=(y+ib_{\rm c}/2)/r^2$, $y_{12}=-ye^{i\alpha}/r$,
$y_{21}=-ye^{-i\alpha}/r$, and $y_{22}=(y+ib_{\rm c}/2)$. The
relations for the $\Pi$-model of a transmission line are obtained by
setting $r=1$ and $\alpha=0$, and in this case one has $y_{12}=y_{21}$
and $y_{11}=y_{22}$.  Accordingly, one can assign a 2$\times2$
admittance matrix to each transmission or transformer line in the
power grid.

In a grid with $N$ nodes, several lines can be connected to a node
$j$, and we define $I_j$ as the total current leaving this node, i.e.
the sum of the currents flowing into the connected lines. Because all
currents and voltages are linearly related, one can write
\begin{equation}
 I = Y V \,,
 \label{eq:i-v}
\end{equation}
where $I=(I_1,\ldots,I_N)^{\rm T}$ and $V=(V_1,\ldots,V_N)^{\rm T}$
are the column vectors of node currents and voltages, respectively,
and $Y$ is an $N\times N$ admittance matrix.
In general, one needs to consider also the effect of leakage
currents at some nodes, which can be taken into account by a change of
the diagonal elements of $Y$ \cite{matpower-manual}.

A further aspect is to consider impedances between the points of power
generation/consumption and the points of power injection/ejection
into/from the grid, as illustrated in figure~\ref{fig02}(b).  This
amounts of extending the grid structure by adding an internal node (of
power generation/consumption) to each node that after this addition
represents a terminal node of power injection/ejection
\cite{Nishikawa/Motter:2015}.  Each internal node is connected by one
line with admittance $y_j$ to the corresponding terminal node $j$ [see
  also figure~\ref{fig02}(b)].\footnote{The steady-state values of
  these admittances can be affected in transient time intervals by the
  armature reaction effect \cite{Machowski/etal:2008}, which will not be
  considered here.}  With the zero current vector and voltage vector
$V$ for the passive terminal nodes in the extended grid, and the
current vector $I'$ and voltage vector $V'$ for the internal nodes,
the relation between voltages and currents becomes
\begin{equation}
\left [\begin{array}{c}
I' \\ 0 \end{array}\right ]
=\left\{
\left [\begin{array}{cc}
Y_{\rm d} & -Y_{\rm d} \\
-Y_{\rm d} & Y_{\rm d}
\end{array}\right ] 
+\left [\begin{array}{cc}
0 & 0 \\
0 & Y
\end{array}\right]\right\}
\left[\begin{array}{c}
V' \\ V  \end{array}\right]
= \left [\begin{array}{cc}
Y_{\rm d} & -Y_{\rm d} \\
 -Y_{\rm d} & Y+Y_{\rm d}
\end{array}\right] 
\left [\begin{array}{c}
V' \\ V \end{array}\right]\,,
\label{eq:i-v-extended}
\end{equation}
where $Y_{\rm d}$ is the diagonal matrix with elements $(Y_{\rm
  d})_{jj}=y_j$.  Eliminating $V$ by using the second line of the
block-matrix equation~(\ref{eq:i-v-extended}), $V=(Y+Y_{\rm
  d})^{-1}Y_{\rm d}V'$, the first line of (\ref{eq:i-v-extended})
becomes
\begin{equation}
 I' = Y' V' \,,
 \label{eq:i-v-reduced}
\end{equation}
where $Y'=Y_{\rm d}-Y_{\rm d}(Y+Y_{\rm d})^{-1}Y_{\rm d}$.  This
reduction to an effective grid consisting of only the $N$ internal
nodes is commonly referred to as Kron reduction.  By Kron reduction, nodes being connected
in the original graph by a path of passive nodes become directly
linked in the reduced graph \cite{Doerfler/Bullo:2013}.  Accordingly,
the graph associated with equation~(\ref{eq:i-v-reduced}) corresponds
to a complete graph between the internal nodes.

In the following, we write all further equations for the grid model
without consideration of the internal nodes, i.e.\ for the relation
(\ref{eq:i-v}) between currents and voltages.  The corresponding
equations for the extended grid model with internal nodes are obtained
by replacing $Y$ with $Y'$ and $V$ with $V'$.

\subsection{Power flow dynamics}
\label{subsec:dynamics}

In the synchronous state of a power grid, all nodal currents and
voltages have a common harmonic time dependence, $I_j(t)={\rm
  Re}\,[I_je^{-i\omega_{\rm r} t}]$, $V_j(t)={\rm Re}\,[V_j
  e^{-i\omega_{\rm r} t}]$, where $I_j=|I_j|e^{i\varphi_j}$ and
$V_j=|V_j|e^{i\theta_j}$ are the complex amplitudes (or ``phasors''),
and $\omega_{\rm r}$ is the common reference frequency of the
generators.  Generators and loads are in first place characterized by
their power supply and consumption. It is thus common to rewrite the
relation~(\ref{eq:i-v}) in terms of powers $S_j=V_j
I_j^\star=P_j+iQ_j$, where $I_j^\star$ denotes the complex conjugate
of $I_j$, and $P_j={\rm Re}\,S_j$ and $Q_j={\rm Im}\,S_j$ are the real
and reactive electrical powers generated by node $j$ ($P_j<0$ if power is consumed).
 Eliminating the currents with the help of
equation~(\ref{eq:i-v}), $S_j=V_j I_j^\star= \sum_k V_j
V_k^\star Y_{jk}^{\star}$, and writing $Y_{jk}=|Y_{jk}| \exp[i
  (\gamma_{jk}+\pi/2)]$,\footnote{The shift of $\pi/2$ in this
  definition of the phase angle $\gamma_{jk}$ is introduced to yield a
  sine function in the subsequent equation~(\ref{eq:swing}), which
  then has the common form of a Kuramoto model of second order
  \cite{Kuramoto1975, Acebron/etal:2005, Rodrigues/etal:2016}.}  one obtains
\begin{eqnarray}
P_j &=& \sum_{k} |V_{j}||V_{k}||Y_{jk}|
\sin\left(\theta_{j}-\theta_{k}-\gamma_{jk}\right),
\label{eq:pf-p}\\
Q_j &=& \sum_{k} |V_{j}||V_{k}||Y_{jk}|
\cos\left(\theta_{j}-\theta_{k}-\gamma_{jk}\right)\,.
\label{eq:pf-q}
\end{eqnarray}
Balancing these electrical powers with the ``mechanical powers"
$S_j^{\rm (m)}=P_j^{\rm (m)}+iQ_j^{\rm (m)}$ determines the steady
state of the power grid.
The corresponding equations are called power flow equations. In power
flow modelling, values $P^{\rm (m)}_j$ and $|V_j|$ are usually given
for generator nodes and values $P^{\rm (m)}_j$ and $Q^{\rm (m)}_j$ for
load nodes.

For describing the mechanical coupling to the electricity grid, we use
the synchronous machine model for both generators and loads with
constant voltage moduli $|V_j|$ \cite{Nishikawa/Motter:2015}. If the
frequency of a synchronous machine deviates from $\omega_{\rm r}$, the
voltage phase angles $\theta_j$ become time-dependent and evolve in
time according to the swing equations \cite{Machowski/etal:2008, Rohden/etal:2012}
\begin{equation}
H_j \ddot{\theta}_j + D_j \dot{\theta}_j = P_j^{\rm (m)}-P_j=
P_j^{\rm (m)}- \sum_k K_{jk}\sin\left(\theta_j-\theta_k-\gamma_{jk}\right)\,.
\label{eq:swing}
\end{equation}
Here, $K_{jk} =|V_j||V_k||Y_{jk}|$ are the coupling constants, and
$H_j$ and $D_j$ are inertia and damping constants, respectively.  The
inertia constant $ H_j $ is connected with the rotating mass of a
motor or conventional generator.  The constants $D_j$ effectively
include both electrodynamical and mechanical damping effects, as well
as primary control measures \cite{ucte}. Because only differences
between phase angles appear in equation (\ref{eq:swing}), the dynamics
are invariant under a constant phase shift
$\theta_j\to\theta_j+\beta$.  For uniqueness, we have set $\theta_1=0$
as a reference angle.

Fixed points of equation~(\ref{eq:swing}) correspond to a stationary
synchronous state described by equations (\ref{eq:pf-p}) and
(\ref{eq:pf-q}), where mechanical and electrical power are balanced
($P_j^{\rm (m)}=P_j$).  In this state, the frequency deviations
$\omega_j=\dot\theta_j$ from the reference frequency $\omega_{\rm r}$
become zero and the phase angles stay constant. Because
equations~(\ref{eq:pf-p}) and (\ref{eq:pf-q}) are non-linear in the
$\theta_j$, there can be more than one fixed point
\cite{Mehta/etal:2015, Delabays/etal:2016}.  As our reference fixed point, we
take the solution of equations~(\ref{eq:pf-p}) and (\ref{eq:pf-q})
obtained by a Newton-Raphson method with starting angles
$(\theta_1,\dots,\theta_N)=(0,\dots,0)$, yielding the fixed point
vector $(0,\theta_2^\ast,\dots,\theta^{\ast}_N)$.  If the
provided/consumed mechanical power $P_j^{\rm (m)}$ deviates from the
electrical power $P_j$, the frequencies
$\omega_j=\dot\theta_j$ are driven away from zero and the phase angles
from their fixed point state.

\subsection{Parameters for extended and simplified variants of 
the IEEE RTS-96 grid}
\label{subsec:parameters}

For our analysis, we use the IEEE RTS-96 grid with structure shown in
figure~\ref{fig:grid-illus}, where for each node
is listed a load contribution $\Pi_j^{\rm (l)}\le0$ and for a subset an additional
generator contribution $\Pi_j^{\rm (g)}>0$. The latter nodes are specified as generator nodes in the RTS-96 grid and 
the other as load nodes. The $N_{\rm g}=33$ generator nodes are numbered 1 to 33 and the
$N_{\rm l}=40$ load nodes
34 to 73. Setting $\Pi_j^{\rm (g)}=0$ for the load nodes, the $P_j^{\rm (m)}$
are fixed by $P_j^{\rm (m)}=\Pi_j^{\rm (g)}+\Pi_j^{\rm (l)}$.
Furthermore, the $|V_j|$ are listed
for generator nodes, and the $Q_j^{\rm (m)}$ for load
nodes. For each lines in the network, the admittances [see
  equation~(\ref{eq:branch-model})], as well as the leakage currents
(or shunt elements) are provided, i.e.\ all information to calculate the elements $Y_{jk}$ of the
admittance matrix in equation (\ref{eq:i-v}).  Altogether this allows one
to determine stationary states based on the power flow equations
(\ref{eq:pf-p}) and (\ref{eq:pf-q}).

For solving the dynamic equations (\ref{eq:swing}), the
damping and inertia constants $D_j$ and $H_j$ are
also needed. Corresponding values are not listed for the IEEE RTS-96
grid (except $H_j$ for generator contributions). To assign values to these 
constants, we use estimates given in
Ref.~\cite{Motter/Nishikawa:2012}, resulting in $D_j$ being all the same, namely
$D_j\cong13$~MWs, and $H_j$ being proportional to the $P_j^{\rm (m)}$
beyond a minimum value,
$H_j=\max(0.064~\mbox{MWs$^2$},2.54\times10^{-4}P_j^{\rm (m)})$.
Moreover,  when taking into account internal nodes, the admittances $y_j$ between
internal and terminal nodes are needed. To specify them, we
use the empirical relation $y_j=92.8(P_j^{\rm (m)})^{-1.3}$ \cite{Motter/Nishikawa:2012}.

As for the units, powers $S_j=P_j+iQ_j$ are expressed in terms of a
power base $P_b=100$~MW.  Times (frequencies) are given in units of
seconds (Hertz).  Accordingly, the units for the coupling constants
$K_{jk}$, the damping constants $D_j$ and the inertia constants $H_j$
in equation~(\ref{eq:swing}) are $P_b=100$~MW, 100~MWs and
100~MWs$^{2}$, respectively.  Voltages are given in a so-called
per-unit system (p.u.), to avoid a modelling of different voltage
levels in the grid.  A voltage base $V_{b} $ is specified as the
nominal value for each voltage level. Admittances $Y_{jk}$ then have
units $Y_b=P_b/V_b^2$.  In the following, all quantities are given in
dimensionless numbers with respect to these units.

Taking these properties altogether, one arrives at a model that
incorporates the internal nodes as well as the heterogeneities both in the nodes, i.e.\ the
different values of $P_j^{\rm (m)}$, $Q_j^{\rm (m)}$ and $|V_j|$, and
in the transmission lines, i.e. the different values of the
admittances $Y_{jk}$.  We refer to this
model as ``Het+IN''.

To investigate the impact of simplifications of the network structure
on the evaluation of grid stability, we consider the following model
variants.  In the first variant, referred to as ``Het-IN'', the
internal nodes are neglected.  As a consequence, there are no passive
nodes and, instead of the admittance matrix $Y^{\prime}$, the
admittance matrix $Y$ is used, see equations~(\ref{eq:i-v}) and (\ref{eq:i-v-reduced}).

\begin{figure*}[t]
\centering
\includegraphics[width=0.9\textwidth]{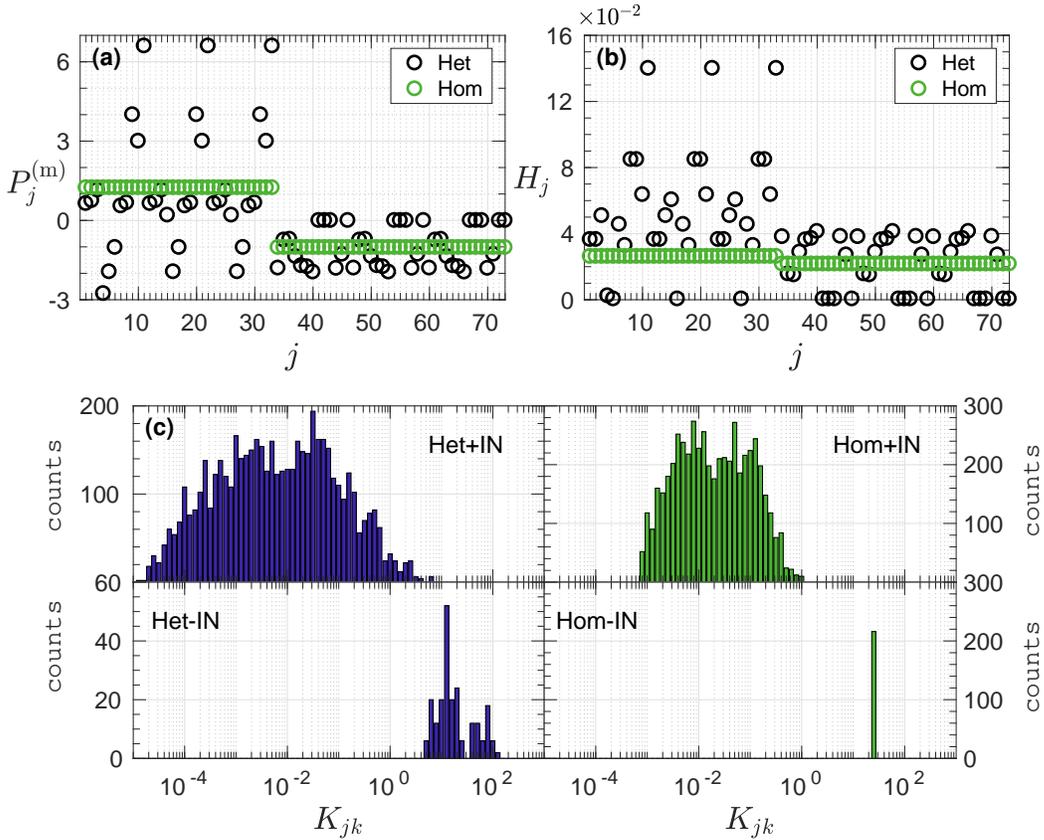}
\caption{\protect
  (a) Mechanical powers $P^{\rm (m)}_j$ for all nodes $ j $ in
  the heterogeneous models (black bullets) and the homogenised counterparts (green bullets).
  (b) Corresponding inertia constants $H_j$ for all nodes.
  (c) Histograms of the coupling constants $K_{jk}$ for each of the four
  models sketched in figure~\ref{fig02}(c) without diagonal elements.}
\label{fig:parameters}
\end{figure*}

Two further variants ``Hom$\pm$IN'' (with and without internal nodes)
are obtained by homogenising the transmission line properties as well
as the generated and consumed powers. Corresponding homogeneous power grid
structures have been investigated in various previous studies \cite{Tamrakar/etal:2018, Filatrella/etal:2008, Lozano/etal:2012, Menck/etal:2014, Witthaut/etal:2012, Jung/Kettemann:2016},
typically for the case of loss-free transmission lines.  For the
homogenisation, we here also consider a loss-free situation, i.e.\ we
neglect the resistances of the transmission lines and the leakage
currents.  We additionally set $r=1$ for the transformer lines
(transformer phase shifts $\alpha$ are zero in the IEEE RTS-96 grid). Then the
arithmetic mean of the line admittances is assigned to all
transmission lines. As for the powers $P_j^{\rm (m)}$, we assign
to the generator nodes their average power ($\bar P=\sum_{j=1}^{N_{\rm g}} P_j^{\rm (m)}/N_{\rm g}$)
and to the load nodes the one 
ensuring a balance between generated and consumed
power ($-\bar  PN_{\rm g}/N_{\rm l}$).\footnote{Let us note that a corresponding balance under
  consideration of losses, i.e.\ when keeping the resistances of the
  lines, can usually be ensured only, if one generator node is allowed
  to exhibit a mechanical power different from the other generator nodes. Accordingly, in that
  case one would not consider a fully homogenised grid.}  Analogously,
we take the arithmetic means for setting the voltage moduli of the
generator nodes and for setting the reactive power of the load nodes.

All four models, Het$\pm$IN and Hom$\pm$IN are illustrated in
figure~\ref{fig02}(c) and the relevant parameters
entering equation (\ref{eq:swing}) are plotted in
figure~\ref{fig:parameters}. For the heterogeneous models, the
$P^{\rm (m)}_j$ [figure~\ref{fig:parameters}(a)] and $H_j$
[figure~\ref{fig:parameters}(b)] cover a quite broad range (black circles), while
in the homogenised models, two different fixed values for generator and load nodes are 
given in each case (green circles).
Let us note that seven of the generator nodes have an ``effective load power'' $P_j^{\rm (m)}<0$,
because $\Pi_j^{\rm (g)}<|\Pi_j^{\rm (l)}|$. Moreover, thirteen of the load nodes have
$P_j^{\rm (m)}=\Pi_j^{\rm (l)}=0$ and therefore act like passive nodes in the
synchronous state of operation, but not in the dynamical case with an overall imbalance of 
mechanical and electrical power. These load nodes are
marked by open circles in figure~\ref{fig:grid-illus}.

In figure~\ref{fig:parameters}(c), histograms of the coupling constants
$K_{jk}$ are shown for the four models. As can be seen from this
figure, the heterogeneous models yield distributions spanning a
broader range, including those obtained for the homogeneous
counterparts.  For the models Het-IN and Hom-IN, the coupling matrix
$K$ resembles the adjacency matrix of the network without internal
nodes, i.e.\ it has non-zero entries only for transmission lines
connecting two nodes. There are in total 108 of these entries for the
IEEE RTS-96 grid.  For the models Het+IN and Hom+IN with internal
nodes, $K$ is a full matrix due to the Kron reduction. This matrix has
$73\times73=5329$ non-zero entries, resulting in the different numbers
of counts in the corresponding histograms in
figure~\ref{fig:parameters}(c). 
Moreover, in the synchronous state, equation~(\ref{eq:pf-p}) must be fulfilled for the same
$P^{\rm (m)}_j$,
implying that the couplings $K_{jk} =|V_j||V_k||Y_{jk}|$ for the models without internal nodes, where 
$Y_{jk}\ne0$ only for nodes connected by transmission lines,
must be in general larger than that of the corresponding models with internal nodes.
Accordingly,  their histograms are centred around coupling constants larger
than that of their counterparts.

Let us note also, that, despite of the homogenising, the histogram for
the Hom$+$IN model exhibits an appreciable width. This is because the
matrix elements of $Y'$ in equation~(\ref{eq:i-v-reduced}) vary due to
the inhomogeneity in the link topology of the grid, which is present
also for the homogenised variants.  Even for the Hom-IN model, the
width is not zero, because the $K_{jk}$ depend also on the voltage moduli, which in the
fixed point state vary slightly for the load nodes.

\section{Basin stability}
\label{sec:basin-stability}
If mechanical and electrical powers are balanced, all phases have the
same angular velocity $\dot{\theta}_{1} = \dot{\theta}_{2} = ... =
\dot{\theta}_{N}=\omega_{\rm r}$ in the fixed point state.  There is
another type of stationary solutions of equation~(\ref{eq:swing}),
which correspond to limit cycles \cite{Menck/etal:2014}, where phase
angles and frequencies of nodes follow closed trajectories. These
limit cycles imply undesirable frequency variations.  Depending on the
amplitude of a perturbation from the fixed point state, the system
either returns to this point or turns into the undesirable state of
asynchronous voltage phase dynamics, which we
refer to as ``grid failure'' in the following.

As a measure for the stability of the synchronous fixed point state
one can take the probability of return to this state after
perturbations $(\Delta\theta,\Delta\omega)$ of single nodes. The
$(\Delta\theta,\Delta\omega)$ are drawn from a probability density
$p_0(\Delta\theta,\Delta\omega)$, which we take to be uniform in the
region $\mathcal{R}=[-\pi,\pi[\times[-2\pi,2\pi]$. After a
    perturbation of node $j$, the probability $S_j$ of the grid to
    return to the fixed point state is then given by
\begin{equation}
 S_j = \int \!\!\mathrm{d}(\Delta\theta)\,
 \mathrm{d}(\Delta\omega)p_0(\Delta\theta,\Delta\omega) \chi_j(\Delta
 \theta, \Delta \omega)\!\! 
 =\frac{1}{8\pi^2}\int_{\mathcal{R}} \!\!\mathrm{d}(\Delta\theta)\,
 \mathrm{d}(\Delta\omega)\chi_j(\Delta
 \theta, \Delta \omega)\,.
\label{EQ:basinStab}
\end{equation}
Here $\chi_j(\Delta \theta, \Delta \omega)$ is an indicator function,
being one if the grid returns to the fixed point after a perturbation
$(\Delta\theta,\Delta\omega)$ of node $j$, and zero otherwise. 
We define the set of points
$\mathcal{B}_j=\{(\Delta\theta,\Delta\omega)|\chi_j(\Delta \theta, \Delta \omega)=1\}$
as the ``attraction basin'' of the fixed point,
where one should keep in mind that we consider single node perturbations and
not the full $(2N)$-dimensional phase space around the fixed point.
For the uniform distribution $p_0(\Delta\theta,\Delta\omega)$, $S_j$ in 
equation~(\ref{EQ:basinStab}) equals
the fraction of return points in $\mathcal{R}$, which has been
referred to as ``basin stability'' \cite{Menck/etal:2013}. The probability of grid failure is $(1-S_j)$.

To determine the basin stabilities $S_j$, we scan the perturbation region $\mathcal{R}$
by using a fine raster grid of spacing
$(\delta\theta,\delta\omega)=(0.1,0.1)$.  For each point
$(\Delta\theta,\Delta\omega)$ in this grid, we solved
equation~(\ref{eq:swing}) for the initial conditions
$(\theta_j,\omega_j)=(\theta_j^\ast+\Delta\theta,\Delta\omega)$ and
$(\theta_k,\omega_k)=(\theta_k^\ast,0)$ for all $k\ne j$.  This is
done numerically by using a Runge-Kutta solver of fourth order.

The grid is considered to have returned to the fixed point
[$\chi_j(\Delta\theta,\Delta\omega)=1$], if there exits a time beyond
which the magnitudes of all frequencies $\omega_k(t)$ remain below a
threshold $\epsilon$, i.e. when $|\omega_k(t)|<\epsilon$ for long
times and all $k$.  Otherwise, the grid is considered to have turned
into a limit cycle [$\chi_j(\Delta\theta,\Delta\omega)=0$]. The
threshold is set as $\epsilon=0.01$, implying that frequency
deviations below $\epsilon$ corresponds to values well below the onset
of primary control measures [typically starting at deviations
  $\omega/(2\pi)$ of about 10~mHz]. In practice, we have taken a total
simulation time of 25~seconds for the analysis of
$\chi_j(\Delta\theta,\Delta\omega)$.  We carefully checked that the
results did not change when increasing the total simulation time.  The
integral in equation (\ref{EQ:basinStab}) is evaluated numerically by
summing over the points of the raster grid.

\begin{figure*}[t!]
\centering
\includegraphics[width=0.95\textwidth]{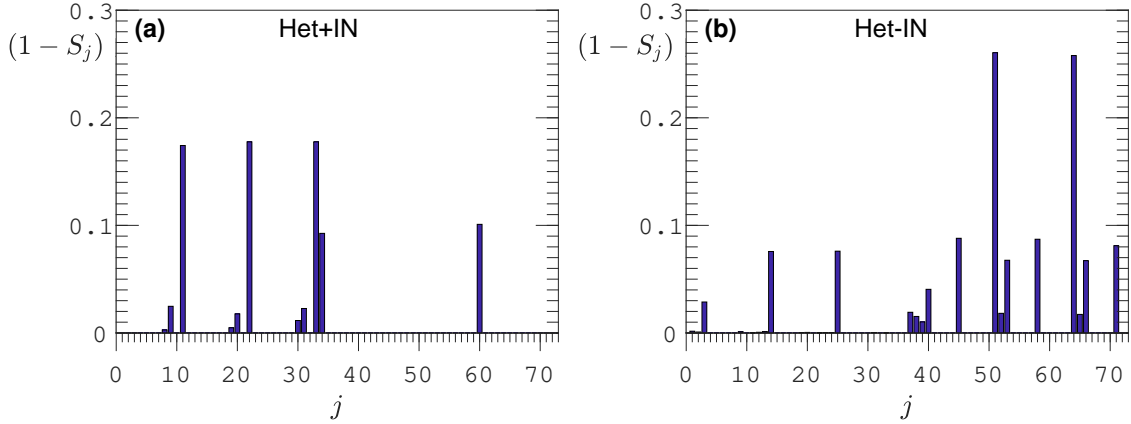}
\caption{\protect Probabilities $(1-S_j)$ of grid failure after perturbing node $j$ 
[see equation~(\ref{EQ:basinStab})] for
models (a) Het+IN and (b) Het-IN.}
\label{fig:basin-stability}
\end{figure*}

Figure~\ref{fig:basin-stability} shows the probability $(1-S_j)$ of
grid failure after a perturbation of node $j$ for models (a) Het+IN
and (b) Het-IN. For the homogeneous models, we always found that the
grid returns to the fixed point state for perturbations in
$\mathcal{R}$, i.e.\ $(1-S_j)=0$ for all $j$. Comparing the results
for the models Het+IN and Het-IN, we see that for the model Het+IN
less perturbed nodes $j$ can lead to grid failure. This can be
understood by the fact that the coupling matrix for the 
Het+IN model represents a complete graph and accordingly a perturbation of
node $j$ becomes rapidly distributed over all other nodes.  For the
Het-IN model, by contrast, node $j$ has only a few nearest-neighbour
nodes (see also figure~\ref{fig:grid-illus}) and a perturbation can have
a stronger impact on the loss of grid stability. The change of the coupling structure
is also reflected in the fact that almost all of
the perturbed nodes in model Het+IN, which can cause grid failure, do
never lead to grid failure in model Het-IN (for perturbation in
$\mathcal{R}$). 

In the Het+IN model, the nodes with high grid failure probabilities $(1-S_j)$
always have relatively high values $|P_j^{\rm (m)}|$. In particular, the
three nodes 11, 22, and 33 with the highest $(1-S_j)$ have also the
largest $|P_j^{\rm (m)}|$. In the Het-IN model this correlation is not
present. Interestingly, there the neighbours 
of the nodes constituting dead ends in the grid topology
exhibit the highest grid failure probabilities,
see the $(1-S_j)$ values in figure~\ref{fig:basin-stability}(b)
for the nodes 51 and 64, which are the neighbours of the
dead-end nodes 14 and 25 (cf.\ figure~\ref{fig:grid-illus}). 
The dead-end nodes themselves belong to the group
of nodes with highest failure probabilities larger than 7\%. 
These findings are in agreement with those reported in \cite{Menck/etal:2014}. There, a 
grid structure corresponding to the homogenised variant without internal nodes
was analysed, but with an about 100 times larger range of initial frequency perturbations 
$\Delta\omega$. 

However, nodes 45, 58, and 71 in figure~\ref{fig:basin-stability}(b) exhibit
also grid probabilities larger than 7\%. 
As can be seen from figure~\ref{fig:grid-illus}, each of these
nodes has only two neighbours and one of them
has high $P_j^{\rm (m)}$ [see also figure~\ref{fig:parameters}(a)]
and accordingly high inertia $H_j\propto P_j^{\rm (m)}$ [see figure~\ref{fig:parameters}(b)].
The high inertia implies that the corresponding nodes react very slowly in response to perturbations
at the neighbouring node, i.e.\ the three nodes 45, 58, and 71 can be considered
as ``effective dead ends''.  That in this case perturbations at the effective dead end nodes themselves, 
rather than their neighbours,
most likely lead to grid failure, can be understood from the fact, that the 
neighbouring nodes with high inertia act like ``fixed'' rather than ``loose''  ends.

Overall, the perturbed nodes that can cause grid
failure in figures~\ref{fig:basin-stability}(a) and (b)
are smaller in number than the ones showing always return to
the synchronous fixed point state.  The maximal
probabilities for grid failure are below 20\% for model Het+IN and
30\% for model Het-IN, implying that for the corresponding perturbed
nodes the attraction basins $\mathcal{B}_j$ cover more than
80\% and 70\% of $\mathcal{R}$, respectively. In the following we will
refer to the nodes with $S_j<1$ as ``basin-escapable'' (BE) nodes.

\section{Return times}
\label{sec:return-times}
Since all single node perturbations for the homogenised models and
most of them for the heterogeneous models lead to return of the power
grid to the stable fixed point state, the grid stability should be
assessed in more detail. In fact, not only the return to a synchronous
state of operation matters, but also the time needed for this return
after a perturbation. Therefore, we introduce a return time $\tilde
t_{jk}(\Delta\theta,\Delta\omega;\epsilon)$ as the minimum time for
$|\omega_k(t)|<\epsilon$ to hold for all $t >\tilde
t_{jk}(\Delta\theta,\Delta\omega;\epsilon)$, after node $j$ is
perturbed with an amplitude $(\Delta\theta,\Delta\omega)$.  In case
the grid state turns into a limit cycle for the perturbation
$(\Delta\theta,\Delta\omega)$, $\tilde
t_{jk}(\Delta\theta,\Delta\omega;\epsilon)$ is assigned infinity for
all $k$. In this way, the $\tilde t_{jk}$ [for all
  $(\Delta\theta,\Delta\omega)$ and all $k$] contain also the information
of $S_j$ in equation~(\ref{EQ:basinStab}). 

\begin{figure*}[t]
\centering
\includegraphics[width=0.8\textwidth]{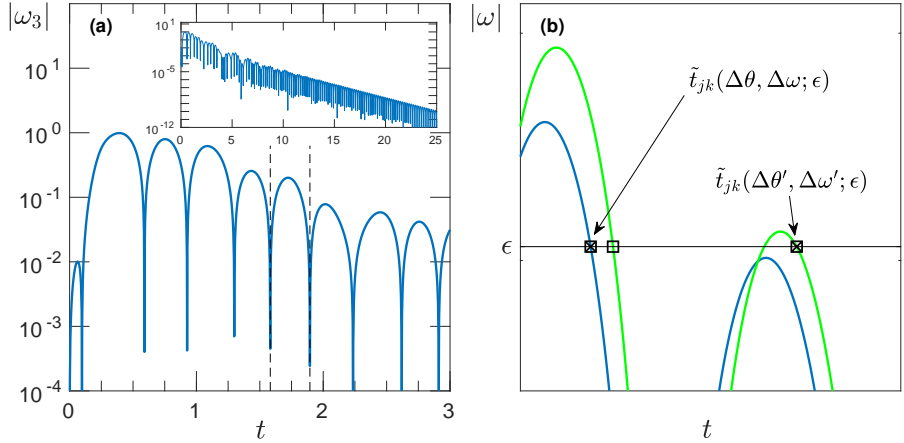}
\caption{\protect
  (a) Part of the trajectory $|\omega_3(t)|$ of node $k=3$
  if node $j=9$ is perturbed 
  with $\Delta\theta=2.5$ and $\Delta\omega=3.1$. The vertical dashed lines
 indicate one oscillation branch between zero crossings of $\omega_3(t)$ (see text) and
 the inset displays the trajectory over the full simulation time.
  (b) Illustration of a jump in the return times $\tilde t_{jk}(\Delta\theta,\Delta\omega;\epsilon)$
  upon changing the initial perturbation
  $(\Delta\theta,\Delta\omega)$ to a nearby $(\Delta\theta',\Delta\omega')$:
  The blue line marks the trajectory of $|\omega_k(t)|$ for $(\Delta\theta,\Delta\omega)$ and the
  green one for $(\Delta\theta',\Delta\omega')$. For the blue line, $|\omega_k(t)|$ remains below
  the threshold $\epsilon$ (horizontal line) after the time $\tilde t_{jk}(\Delta\theta,\Delta\omega;\epsilon)$ 
  (point marked by the crossed square). For the green line, the corresponding intersection
 of $|\omega_k(t)|$ with the horizontal line in the respective oscillation branch (point marked by the open square) does not give the correct return time. This is because in the successive oscillation branch the local maximum of the trajectory exceeds $\epsilon$. The correct return time for the green line 
 thus is $\tilde t_{jk}(\Delta\theta',\Delta\omega';\epsilon)$ as indicated in the figure 
 (point marked by the crossed square). Due to the change of the oscillation branch,
 $\tilde t_{jk}(\Delta\theta',\Delta\omega';\epsilon)$ is not close to 
 $\tilde t_{jk}(\Delta\theta,\Delta\omega;\epsilon)$.}
\label{fig:return-time-illus}
\end{figure*}

Figure~\ref{fig:return-time-illus}(a) shows a part of a representative
trajectory of $|\omega_3(t)|$ if node 9 is perturbed ($j=9$, $k=3$)
with $(\Delta\theta,\Delta\omega)=(2.5,3.1)$; the full trajectory over the entire simulation
time is displayed in the inset. Similar
trajectories are obtained for other node pairs. As can be seen from
this figure, $|\omega_3(t)|$ shows oscillations and one can
subdivide the trajectory into oscillation branches between successive
zero crossings of $\omega_3(t)$. These zero crossings correspond to the minima (``cusps'')
in the semi-logarithmic representation of $|\omega_3(t)|$ in figure~\ref{fig:return-time-illus}(a).
The exact value $(-\infty)$ of these minima is, however, not fully approached due to
the finite time iteration step in the numerics. 
For illustration, we have indicated one oscillation branch by the dashed vertical lines
in figure~\ref{fig:return-time-illus}(a). 
  
The appearance of
the oscillation branches can lead to jumps in the functions 
$\tilde t_{jk}(\Delta\theta,\Delta\omega;\epsilon)$,
because, if changing $(\Delta\theta,\Delta\omega)$ slightly, the
crossing of the threshold $\epsilon$ can occur in two different
oscillation branches.  This effect is demonstrated in
figure~\ref{fig:return-time-illus}(b).
For smoothing the dependence of the return times on the initial
perturbations $(\Delta\theta,\Delta\omega)$, we average the $\tilde
t_{jk}$ over a number of $(2n+1)$ equidistantly spaced thresholds
around a reference value $\epsilon$, yielding
\begin{equation}
t_{jk}(\Delta\theta,\Delta\omega)=
\frac{1}{2n+1}\sum_{m=-n}^n
\tilde t_{jk}(\Delta\theta,\Delta\omega;\epsilon+m\Delta\epsilon)\,.
\label{eq:tjk}
\end{equation}
For $\epsilon$ we take the
same value $\epsilon=0.01$ as in the previous section for the
stability basin analysis and
set $n=3$ and $\Delta\epsilon=1.4\times10^{-3}$.\footnote{If one would
be interested only in the grid return times (see below), an
alternative way to obtain smoothly varying times is to consider the
energy $E= \sum_j \frac{1}{2} H_j \omega_j^{2} +
\sum_j \sum_k \frac{1}{4} K_{jk} \cos(\theta_j^{\ast} -
\theta_k^{\ast}-\gamma_{jk}) \left[ (\theta_j-\theta_j^{\ast}) -
(\theta_k-\theta_k^{\ast}) \right]^{2} $ of coupled harmonic
oscillators, which characterises the system in the neighbourhood of
the fixed point, see equation~(\ref{eq:swing}). If the phase space trajectory $\{(\theta_j(t),\omega_j(t))\}$ 
enters the nearest neighbourhood, this energy
 decreases monotonically with time $t$.}

\begin{figure*}[t]
\centering
\hspace*{-0.7cm}\includegraphics[width=1.13\textwidth]{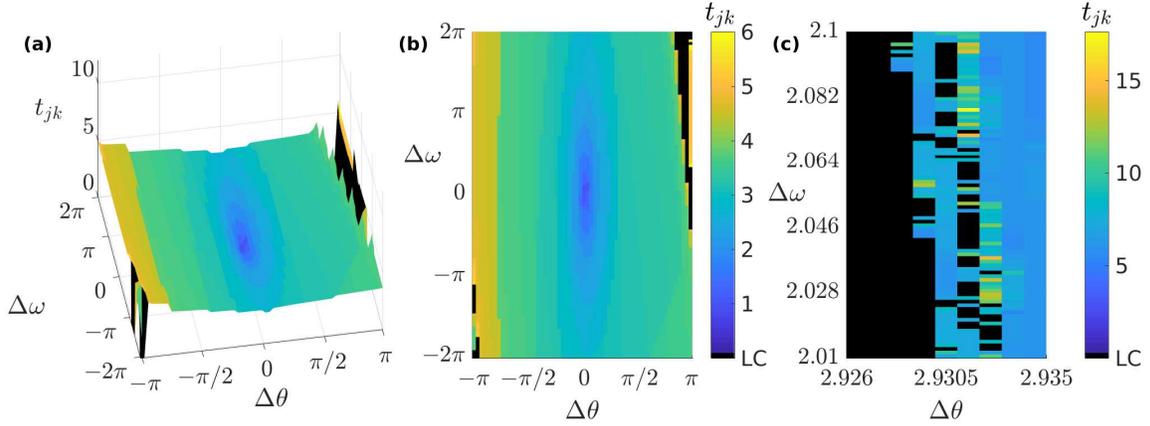}
\caption{\protect Return times $t_{jk}(\Delta\theta,\Delta\omega)$
of node $k=3$ after initial perturbations $(\Delta\theta,\Delta\omega)$ of node $j=9$,
(a) in a surface plot and (b) in a colour-coded 
2d-density plot representation. (c) Part of the 2d-density plot 
scanned with a finer raster grid resolution showing a chaotic behaviour in the transitory region 
between fixed point and limit cycle.}
\label{fig:tjk-illus}
\end{figure*}

As an example, we show in
figure~\ref{fig:tjk-illus} the return times $t_{jk}(\Delta\theta,\Delta\omega)$
of a node $k$ after a perturbation of another node $j$ in the Het+IN model,
(a) in a surface plot and (b) in a colour-coded 
2d-density plot. In the attraction basin $\mathcal{B}_j$, 
the $t_{jk}(\Delta\theta,\Delta\omega)$ become longer with
increasing distance from the fixed point, e.g.\ the Euclidean distance $(\Delta\theta^2+\Delta\omega^2)^{1/2}$.
Points in the perturbation region $\mathcal{R}$ leading to grid failure
are marked black. Far from the fixed point, they typically form a connected domain.
Note that the coloured stripe close to $\Delta\theta=\pi$ at the right side
in figure~\ref{fig:tjk-illus}(b) is connected to the corresponding part at the left side
close to $\Delta\theta=-\pi$. However, in the
transitory region between this domain and the core of the attraction basin, 
a closer inspection with a finer resolution $(\delta\theta,\delta\omega)$ of the raster grid 
reveals that the grid stability becomes very sensitive to the initial perturbation, see figure
figure~\ref{fig:tjk-illus}(c). This
reflects a kind of chaotic behaviour, i.e.\ nearby points $(\Delta\theta,\Delta\omega)$ 
can give different  $\chi_j(\Delta\theta,\Delta\omega)\in\{0,1\}$. For points with
$\chi_j(\Delta\theta,\Delta\omega)=1$ (return to the fixed point) in the transitory region, the 
$t_{jk}(\Delta\theta,\Delta\omega)$ can change rapidly between small and large values.

\subsection{Mean grid and node return times}
\label{subsec:mean-return-times}
The $t_{jk}(\Delta\theta,\Delta\omega)$ contain detailed information on local return
times, including the basin stability $S_j$  of the perturbed node $j$. The latter quantifies how disturbing a node $j$
is for the overall grid stability. For perturbations belonging to the attraction basins $\mathcal{B}_j$,
the $t_{jk}(\Delta\theta,\Delta\omega)$ further allow us
to quantify (i) how disturbing a node $j$ is for sufficiently fast return to the fixed point, 
and (ii) how susceptible a node $k$ is to a perturbation of any node
in the grid (including the node $k$ itself). To this end, we introduce two quantities. The first is the grid return time
$t_j^{\,\rm\scriptscriptstyle grid}(\Delta\theta,\Delta\omega)$, which is the maximum
return time of all nodes $k$ under a perturbation $(\Delta\theta,\Delta\omega)$ of node $j$, that means it is
the time for the entire grid to return to the state of synchronous operation, 
\begin{equation}
t_j^{\,\rm\scriptscriptstyle grid}(\Delta\theta,\Delta\omega)
= \underset{k}{\max} \left\lbrace t_{jk}(\Delta \theta, \Delta \omega)|(\Delta \theta, \Delta \omega)\in\mathcal{B}_j
\right\rbrace.
\label{eq:tgrid}
\end{equation}
The second quantity is the node return time $t_k^{\,\rm\scriptscriptstyle node}(\Delta\theta,\Delta\omega)$.
This is defined as the maximum
return time of node $k$ under a perturbation $(\Delta\theta,\Delta\omega)$ of any of the nodes $j$,
\begin{equation}
t_k^{\,\rm\scriptscriptstyle node}(\Delta\theta,\Delta\omega)
= \underset{j}{\max} \left\lbrace t_{jk}(\Delta \theta,\Delta \omega)|(\Delta \theta, \Delta \omega)\in\mathcal{B}_j
\right\rbrace\,.
\label{eq:tnode}
\end{equation}

\begin{figure*}[t]
\centering
\includegraphics[width=0.8\textwidth]{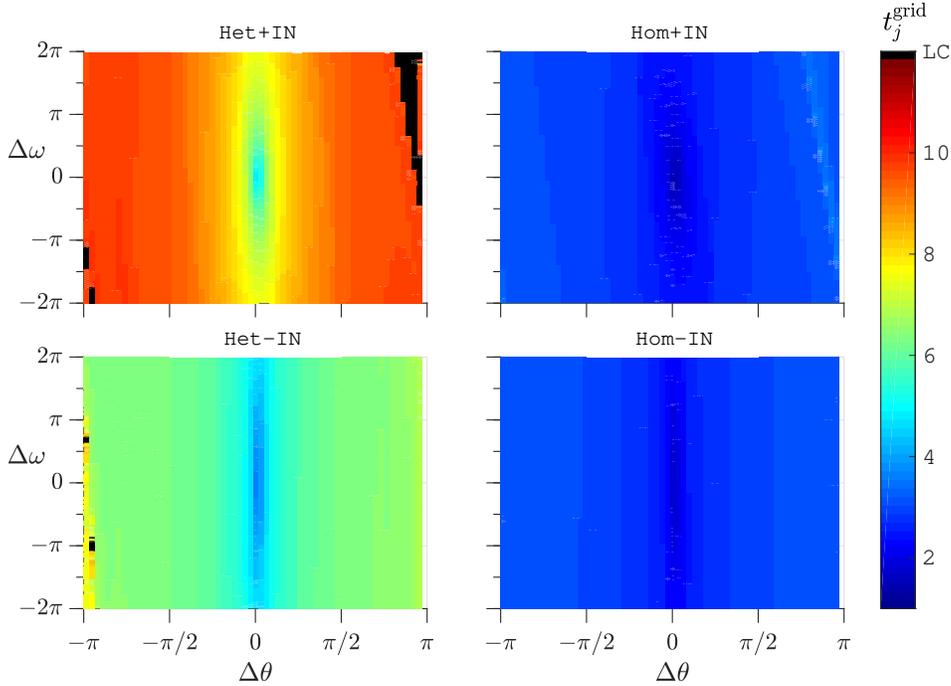}
\caption{\protect
  Surfaces of grid return times $t_j^{\,\rm\scriptscriptstyle grid}(\Delta\theta,\Delta\omega)$
  [equation~(\ref{eq:tgrid})] for $j=9$ and the four model variants in a density plot representation.
  Initial perturbations $(\Delta\theta,\Delta\omega)$ in the heterogeneous models leading
  to grid failure are marked black.}
\label{fig:grid-time-surfaces}
\end{figure*}

Figure~\ref{fig:grid-time-surfaces} shows an example for grid 
return times $t_j^{\,\rm\scriptscriptstyle grid}(\Delta\theta,\Delta\omega)$
for the same perturbed node and the four model variants. 
Strong differences can be seen between the heterogeneous models Het$\pm$IN and
their homogenised counterparts Hom$\pm$IN, and smaller differences between the 
models with and without internal nodes. The typical 
grid return times for the Het$\pm$IN models vary from 1 to 12 seconds 
and are significantly larger than for the Hom$\pm$IN models showing variations between 1 and 5 seconds. As already mentioned above, in the Het$\pm$IN models, 
certain initial perturbations $(\Delta\theta,\Delta\omega)$ lead to grid failure 
(points marked in black), while for the Hom$\pm$IN models the grid returns to the state of synchronous operation
for all initial perturbations  $(\Delta\theta,\Delta\omega)\in\mathcal{R}$.
The main effect of including the internal nodes for both the heterogeneous and homogenized
models shows up in a weaker dependence of the times on the initial frequency perturbation
$\Delta\omega$, as it is reflected in the more stripe-like pattern seen for the Het-IN and Hom-IN models 
compared to the Het+IN and Hom+IN models. 
Furthermore, we note that the transitory region of chaotic-like
return time variations, as discussed in connection with figure~\ref{fig:tjk-illus}, is typically
more pronounced and extended for the Het-IN model. As for the node 
return times $t_k^{\,\rm\scriptscriptstyle node}(\Delta\theta,\Delta\omega)$, a similar
behaviour is obtained, with the same typical differences between the heterogeneous and homogeneous models
and between the variants with and without internal nodes.

To characterise the nodes independent of the initial perturbations, an obvious way is to
introduce the grid and node return times maximised over all $(\Delta \theta,\Delta \omega)$
in the respective attraction basins:
\begin{eqnarray}
T_j^{\rm\scriptscriptstyle grid}&=& 
\max_{(\Delta \theta,\Delta \omega)\in\mathcal{B}_j}\left\lbrace t_j^{\,\rm\scriptscriptstyle grid}(\Delta\theta,\Delta\omega)
\right\rbrace \,,
\label{eq:max-tgrid}
\\
T_k^{\,\rm\scriptscriptstyle node}&=& \underset{(\Delta \theta,\Delta \omega)\in\mathcal{B}}{\max}
\left\lbrace t_k^{\,\rm\scriptscriptstyle node}(\Delta\theta,\Delta\omega)
\right\rbrace \,.
\label{eq:max-tnode}
\end{eqnarray}
Here, $\mathcal{B}=\cap_j\mathcal{B}_j$ is the attraction basin for all possible single-node
perturbations in $\mathcal{R}$ irrespective of $(\Delta \theta,\Delta \omega)$ {\it and} the perturbed node $j$. 
In the Hom$\pm$IN models, we find $T_j^{\rm\scriptscriptstyle grid}\simeq4$ for all nodes,  while in the
Het$\pm$IN models strong variations of $T_j^{\rm\scriptscriptstyle grid}$ from node to node occur. For the
BE nodes, i.e.\ the nodes that can cause grid failure,
these times can be as large as 20. However, the determination of the $T_j^{\rm\scriptscriptstyle grid}$ 
for these
BE nodes (and all $T_k^{\rm\scriptscriptstyle node}$)
needs to be considered with caution,
because their values were seen to depend on the resolution of the raster grid. 
This effect is associated with the chaotic-like dependence of the 
$t_{jk}(\Delta\theta,\Delta\omega)$ on $(\Delta\theta,\Delta\omega)$ 
in the transitory regions between the limit-cycle
domains and the cores of the attraction basins that we discussed above in connection with 
figure~\ref{fig:tjk-illus}. Even larger values for the $T_j^{\rm\scriptscriptstyle grid}$
may be obtained when increasing
the resolution beyond the finest one $(\delta\theta,\delta\omega)=(0.1,0.1)$
used in our analysis. 

\begin{figure*}[t!]
\centering
\includegraphics[width=\textwidth]{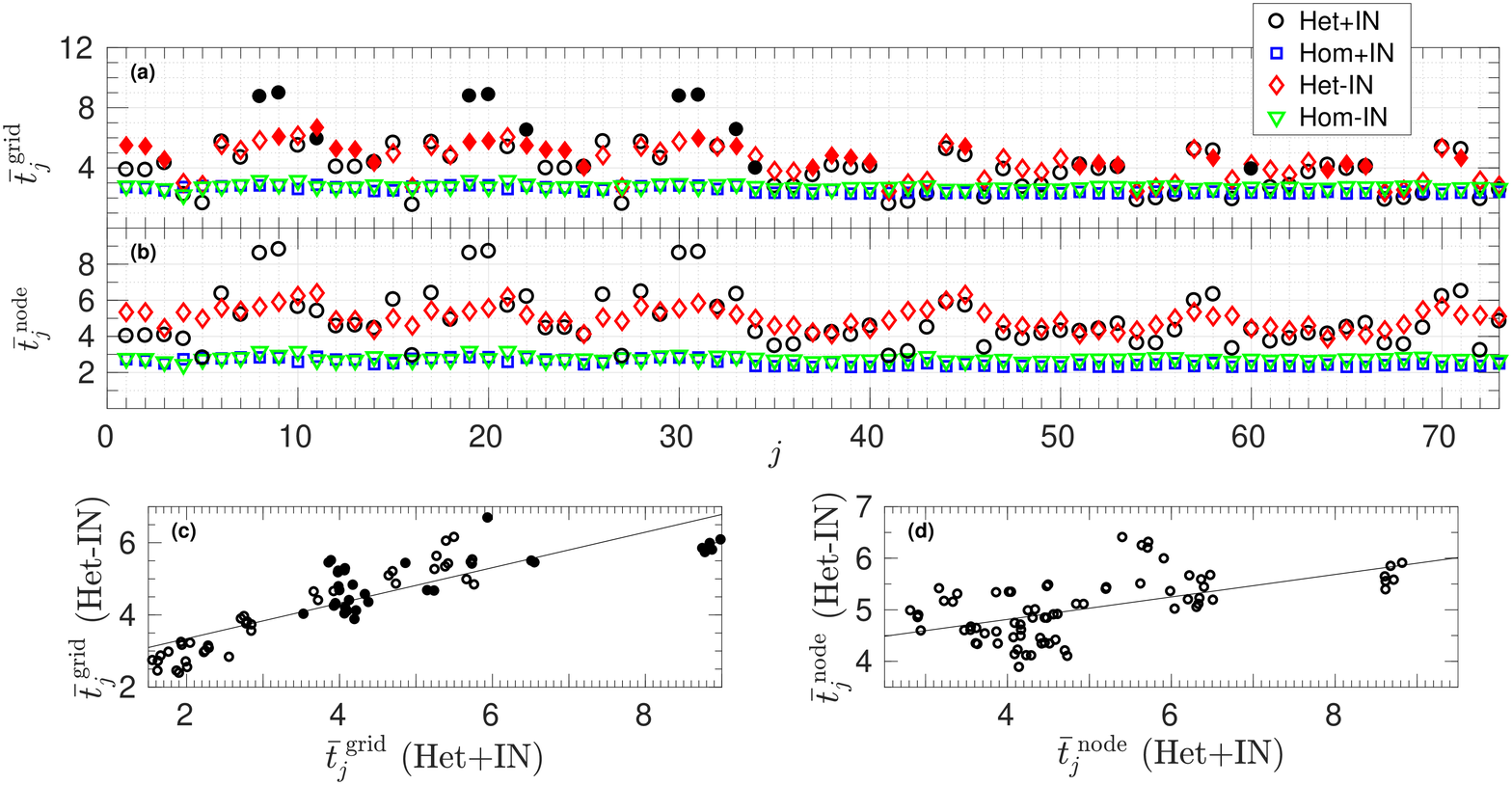}
\caption{\protect 
(a) Mean grid return times and (b) mean node return times 
for each node $j$ and the four model variants. 
A correlation
between the mean return times of the Het+IN and Het-IN models is
tested in scatter plots,
(c) for the grid return times, and (d) for the node return times.
Lines in (c) and (d) correspond to least square fits and
the correlation coefficients are $R\cong0.85$ and $R\cong0.56$ respectively.
In parts (a) and (c) the BE nodes are marked by full symbols.}
\label{fig:grid-times}
\end{figure*}

To ensure robustness of our findings with respect to the raster grid resolution,
we introduce the mean grid and node return times
\begin{eqnarray}
\bar t_j^{\,\rm\scriptscriptstyle grid}
&=&
\frac{1}{N_{\mathcal{B}_j}}
\int_{\mathcal{B}_j} \!\!\mathrm{d}(\Delta\theta)\,\mathrm{d}(\Delta\omega)\,
t_j^{\,\rm\scriptscriptstyle grid}(\Delta\theta,\Delta\omega)\,,
\label{eq:mean-tgrid}\\
\bar t_k^{\,\rm\scriptscriptstyle node}
&=&
\frac{1}{N_\mathcal{B}}
\int_\mathcal{B}\!\!\mathrm{d}(\Delta\theta)\,\mathrm{d}(\Delta\omega)\,
t_k^{\,\rm\scriptscriptstyle node}(\Delta\theta,\Delta\omega)\,,
\label{eq:mean-tnode}
\end{eqnarray}
for an alternative characterisation of the nodes independent of the initial perturbation.
Here, $N_{\mathcal{B}_j}\!=\!\int_{\mathcal{B}_j}\mathrm{d}(\Delta\theta)\,\mathrm{d}(\Delta\omega)\!=\!
8\pi^2 S_j$ and $N_\mathcal{B}\!=\!\int_\mathcal{B}\mathrm{d}(\Delta\theta)\,\mathrm{d}(\Delta\omega)$
are normalisation constants. Indeed, the times $\bar t_j^{\,\rm\scriptscriptstyle grid}$ and  $\bar t_j^{\,\rm\scriptscriptstyle node}$ were found 
to approach constant values with increasing raster grid resolution and
their values converged well for the resolution used. 

Figures~\ref{fig:grid-times}(a) and (b) show a comparison of the 
$\bar t_j^{\,\rm\scriptscriptstyle grid}$ and
$\bar t_j^{\,\rm\scriptscriptstyle node}$
for the four model variants. In case of the homogenised grids, we find 
$\bar t_j^{\,\rm\scriptscriptstyle grid}\simeq3$ and 
$\bar t_j^{\,\rm\scriptscriptstyle node}\simeq2.5$
for all nodes. In the heterogeneous cases, by contrast,
the $\bar t_j^{\,\rm\scriptscriptstyle grid}$ and $\bar t_j^{\,\rm\scriptscriptstyle node}$
vary strongly between different nodes and the times are in general significantly larger
(with a few exceptions for the $\bar t_j^{\,\rm\scriptscriptstyle grid}$). 

As discussed above, perturbation of certain nodes in the heterogeneous models 
can lead to grid failure, and  for the $\bar t_j^{\,\rm\scriptscriptstyle grid}$ it makes sense
to distinguish between these BE nodes and the other ones. 
Therefore we have marked the BE nodes in 
figures~\ref{fig:grid-times}(a) and (c) by full symbols.
Let us remind that, according to 
equations~(\ref{eq:max-tgrid}) and (\ref{eq:mean-tgrid}), the times are
determined with respect to the attraction basins $\mathcal{B}_j$, i.e.\ without including
initial perturbations causing grid failure. As can be seen from figure~\ref{fig:grid-times}(a), 
comparatively large $\bar t_j^{\,\rm\scriptscriptstyle grid}$ 
correspond to perturbations of the BE nodes.  

\begin{figure*}[t!]
\centering
\includegraphics[width=0.5\textwidth]{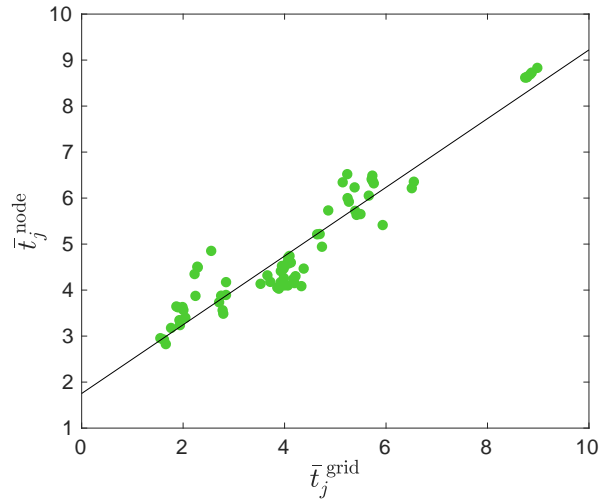}
\caption{\protect Scatter plot of the mean node return times $\bar t_j^{\,\rm\scriptscriptstyle node}$ 
versus the mean grid return times $\bar t_j^{\,\rm\scriptscriptstyle grid}$ for the model Het+IN.
The line corresponds to a least square fit and the correlation coefficient is $R\cong0.95$.}
\label{fig:tgrid-tnode}
\end{figure*}

A remarkable feature can be seen in the Het+IN model.
For this model, the six particular large $\bar t_j^{\,\rm\scriptscriptstyle grid}$ in figure~\ref{fig:grid-times}(a)
are caused by perturbations of  a node in the pairs $(8,9)$, $(19,20)$, and $(30,31)$.
The nodes of these pairs are directly connected by a transmission line, see figure~\ref{fig:grid-illus}. 
Revisiting figure~\ref{fig:basin-stability}, we see that the grid failure 
probabilities of them are not large. Thus, in the Het+IN model we find 
nodes that rarely lead to grid failure but
to long grid return times. Inspection of the environment of the respective node pairs in figure~\ref{fig:grid-illus}
reveals a characteristic pattern: One of the nodes of each pair (nodes 8, 19, and 30)
is strongly coupled to a (load) node with small $|P_j^{\rm (m)}|$ (and hence small
$H_j$, see figure~\ref{fig:parameters}), and it has an
appreciable additional coupling only to its pair partner.
The latter can have several links to
other nodes, but the coupling to its partner node is the dominant one. Accordingly,
a perturbation of it becomes predominantly mediated to its partner.

Looking at the range of the $\bar t_j^{\,\rm\scriptscriptstyle grid}$ in 
figure~\ref{fig:grid-times}(a), we obtain a
slightly wider range from 1.5 to 9 for the Het+IN model compared to the range from 2.5 to 7 for the Het-IN variant.
This shrinking of the range when neglecting internal nodes is even stronger
for the $\bar t_j^{\,\rm\scriptscriptstyle node}$, where we find
a range from 3 to 9 for the Het+IN model and from 4 to 6.5 for the Het-IN variant, see
figure~\ref{fig:grid-times}(b).
While there is no one-to-one correspondence with respect to the rank order of the
return times $\bar t_j^{\,\rm\scriptscriptstyle grid}$ for the Het-IN and Het+IN model,
large times $\bar t_j^{\,\rm\scriptscriptstyle grid}$ in the Het-IN model tend to be
large also in the Het+IN model. A weak linear correlation is demonstrated in figure~\ref{fig:grid-times}(c),
with a (Pearson) correlation coefficient of $R\cong0.85$. In contrast, almost no correlation ($R\cong0.56$)
between the node return times of the Het+IN and Het-IN models is found, see figure~\ref{fig:grid-times}(d).

Finally, we analyse whether more disturbing nodes for the grid 
are also more susceptible for getting disturbed. Indeed, a scatter plot of 
$\bar t_j^{\,\rm\scriptscriptstyle grid}$ versus $\bar t_j^{\,\rm\scriptscriptstyle node}$
reveals a linear correlation in the Het+IN model, see figure~\ref{fig:tgrid-tnode}. The corresponding
correlation coefficient is $R\cong0.95$. Surprisingly, this feature is not seen in the Het-IN model.
A linear correlation between $\bar t_j^{\,\rm\scriptscriptstyle grid}$ and $\bar t_j^{\,\rm\scriptscriptstyle node}$
is also seen in the homogenised models, but not of relevance
in view the small variations of these times.

\section{Summary and conclusions}
\label{sec:conclusions}

For investigating the stability of power grids against single-node perturbations we
have performed extensive numerical calculations for the IEEE RTS-96 grid.
These were carried out for perturbations of the frequencies and voltage phase angles using
the non-linear swing equations, as well as a treatment of generators and loads
based on a synchronous machine approach.
Four model variants were studied to analyse impacts of simplifications, which
we referred to as Hom$\pm$IN and Het$\pm$IN according to 
homogenising grid properties and/or neglecting reactances between terminal and internal nodes. 
The stability of the grid after a perturbations was quantified by the probabilities of return 
to the fixed point state of synchronous operation (basin stability) and additionally by the times
needed for return to this state. These return times allow for a characterisation
of the nodes within the attraction basin of the fixed point, and they include also
a measure for the susceptibility of a node towards a perturbation at another node.
We introduced grid return times for specifying the perturbation strengths of nodes
and node return times for specifying the susceptibility of nodes against perturbations.

The main results for the basin stability can be summarised as follows. 
When homogenising node and transmission line properties, the grid always returns
to its synchronous state. This robustness occurs despite of the large perturbation region considered,
which covers all possible voltage phase angles and frequency deviations of up to 1~Hz.
In the heterogeneous models, perturbation of a couple of nodes can turn the system into a 
stationary state of grid failure with an asynchronous voltage phase dynamics.
When taking into account the internal nodes, the overall grid stability increases.  
This may be expected from the effectively fully connected grid structure in this case, 
which should result in a more mean-field like behaviour with a weakening of heterogeneities. 

Most perturbative nodes with highest probabilities of causing grid failure are different in the
Het-IN and Het+IN model. In the Het-IN variant, these nodes can be correlated with
topological features, namely they are part of dead ends. This feature has been earlier reported
in \cite{Menck/etal:2014} for a grid structure resembling the Hom-IN model, if allowing for
extremely large frequency  perturbations. For frequency perturbations of up to 1~Hz considered in this work, the correlation is seen only in the Het-IN model but not in the corresponding
Hom-IN variant. Our results for the basin stability furthermore suggest that nodes being part of effective
dead ends exhibit a large probability of causing grid failure when perturbed. The effective dead ends
are associated with nodes of high inertia and accordingly slow response. In the Het+IN model,
we found the mechanical power to be the dominant factor controlling basin stability.

Our results for the return times show that the largest times
are about five seconds for the homogenised models. This is small compared to times scales of 
secondary control measures starting at ten seconds \cite{Machowski/etal:2008, ucte}. In contrast, return times for the heterogeneous models
can be 20~seconds and larger. While grid and node return times in the homogenised models are almost the same
for all nodes, a large spread is seen between different nodes in both the Het-IN and Het+IN models. 
This spread turned out to be slightly larger in the Het+IN compared to the Het-IN model, which is 
a bit surprising in view of the overall higher basin stability of the Het+IN model. In the Het+IN model
a linear correlation between grid and node return times is found, meaning that nodes with
higher perturbative capability have higher perturbation receptiveness also. This feature is not
seen in the Het-IN variant.

We conclude that by homogenising node and transmission line properties in a power grid, 
while keeping its link topology, leads to a strong overestimation of grid stability. 
A neglect of reactances between points of power generation/consumption and points of
power injection/ejection can cause wrong identification of most perturbing nodes in the grid
and an underestimation of grid stability. It moreover influences the spread of return 
times as well as correlations between them.

\ack

We thank O.~Kamps, J.~Peinke, and K.~Schmietendorf for very valuable discussions
and gratefully acknowledge funding by the Deutsche Forschungsgemeinschaft (MA 1636/9-1).


\section*{References}

\providecommand{\newblock}{}

\end{document}